\documentstyle[12pt,twoside,amssymbols]{article}
\begin{document}

\def\rg{{\rangle}}
\def\lg{{\langle}}
\def\ra{{\rightarrow}}
\def\v{{\,|\,}}

\def\np{\sim\!p}
\def\nq{\sim\!q}

\def\A{{\cal A}}
\def\B{{\cal B}}
\def\BT{{\tilde{\cal B}}}
\def\C{{\cal C}}
\def\D{{\cal D}}
\def\E{{\cal E}}
\def\F{{\cal F}}
\def\G{{\cal G}}
\def\H{{\cal H}}
\def\HT{{\tilde{\cal H}}}
\def\P{{\cal P}}
\def\Q{{\cal Q}}
\def\W{{\cal W}}
\def\X{{\cal X}}
\def\Z{{\cal Z}}

\def\At{{\tilde A}}
\def\Ct{{\tilde C}}
\def\Dt{{\tilde D}}
\def\Ft{{\tilde F}}
\def\Mt{{\tilde M}}
\def\Pt{{\tilde P}}
\def\Qt{{\tilde Q}}
\def\Rt{{\tilde R}}

\title{Consistent Quantum Reasoning}

\author{Robert B. Griffiths\thanks{Electronic mail: rgrif+@cmu.edu}\\
Department of Physics\\
Carnegie Mellon University\\
Pittsburgh, PA 15213, U.S.A.}

\date{Version of 17 May 1995}
\maketitle

\begin{abstract}
	Precise rules are developed in order to formalize the reasoning
processes involved in standard non-relativistic quantum mechanics, with the
help of analogies from classical physics.  A classical or quantum description
of a mechanical system involves a {\it framework}, often chosen implicitly, and
a {\it statement} or assertion about the system which is either true or false
within the framework with which it is associated.  Quantum descriptions are no
less ``objective'' than their classical counterparts, but differ from the
latter in the following respects: (i) The framework employs a Hilbert space
rather than a classical phase space. (ii) The rules for constructing meaningful
statements require that the associated projectors commute with each other and,
in the case of time-dependent quantum histories, that consistency conditions be
satisfied. (iii) There are incompatible frameworks which cannot be combined,
either in constructing descriptions or in making logical inferences about them,
even though any one of these frameworks may be used separately for describing a
particular physical system.

	A new type of ``generalized history'' is introduced which extends
previous proposals by Omn\`es, and Gell-Mann and Hartle, and a corresponding
consistency condition which does not involve density matrices or single out a
direction of time.  Applications which illustrate the formalism include:
measurements of spin, two-slit diffraction, and the emergence of the classical
world from a fully quantum description.
\end{abstract}


		\section{Introduction}

	Seventy years after non-relativistic quantum mechanics reached what is
essentially its present form, the ongoing controversy over its conceptual
foundations and the proper interpretation of wave functions, measurements,
quantum probabilities, and the like shows no sign of abating.  Indeed, modern
experiments involving neutron diffraction, ions in traps, and quantum optics
seem to cry out for a more satisfactory formulation of basic quantum ideas than
was available to Feynman \cite{fn64} when he remarked, in comparing special
relativity and quantum theory, that while the former was well understood,
``nobody understands quantum mechanics.''

	This lack of understanding manifests itself in various ways.  Take, for
example, the well-known fact that the predictions of standard quantum mechanics
violate Bell's inequality \cite{bl64}.  From this it might seem reasonable to
conclude that all derivations of Bell's inequality contain one or more errors,
in the sense that either the premises or the procedures of logical inference
entering the argument violate one or more principles of standard quantum
mechanics.  Had someone in 1964 attempted to publish an inequality violating
the predictions of special relativity, the experts would have pointed out
precisely where the reasoning went astray as soon as the argument reached the
printed page, if not earlier.  But this did not happen in the case of Bell's
work.  Indeed, much of the subsequent literature on the subject \cite{lit1}
would lead one to believe that standard quantum mechanics is either wrong, or
illogical, or unintelligible, or at the very least contains a hidden assumption
that the world is non-local, an assumption which completely escaped the
physicists who first developed the theory (because they did not understand what
they were doing?).  This despite the fact that all predictions of standard
quantum mechanics seem amply supported by every experimental test designed to
uncover some flaw.

	As a second example, consider the recent assertion by Englert et
al. \cite{ess92} that Bohmian mechanics \cite{bom52,bhi93} makes
predictions (or
at least retrodictions) which disagree with the results of standard quantum
mechanics and also with common sense: detectors designed to detect particles
passing through them can actually be triggered by particles which, according to
the Bohmian interpretation, never come close to the detector.  One might have
imagined that this observation would have prompted advocates of Bohmian
mechanics to withdraw or modify their claim \cite{dgz92} that this theory
reproduces all the results of standard quantum mechanics, especially given an
independent verification \cite{dhs93} of the essential correctness of the
calculations by Englert et al.  Instead, the response \cite{dhs93,dfg93} has
been that standard quantum mechanics, in contrast to Bohmian mechanics, does
not provide an adequate theoretical framework for sensibly discussing whether
the particle passed through the detector; thus one must take the Bohmian result
seriously, as the precise outcome of a well-defined theory, however
counterintuitive it may appear to be.

	The ultimate goal of the research reported in the present article is to
place non-relativistic quantum mechanics on as firm and precise a conceptual
foundation as that of special relativity.  This is not as ambitious a project
as might at first seem to be the case, for two reasons.  First, a large part of
non-relativistic quantum theory is already in a quite satisfactory state, at
least by the somewhat lax standards of theoretical physics. The basic
conceptual difficulties are focused in a small (but critical) area, the point
at which the mathematical formalism of Schr\"odinger's equation, Hilbert space
operators, and the like, which by now is quite well understood, is used to
generate the probabilities needed to compare theoretical predictions with
experimental results. Second, a set of ideas which seem to be adequate for
integrating the probabilistic and deterministic aspects of quantum theory into
a coherent whole are already in the published literature, although their
significance has not been widely appreciated.  The first of these ideas is von
Neumann's formulation of quantum theory using closed subspaces of Hilbert space
to represent quantum properties \cite{vn55a}---to be carefully distinguished
from his theory of measurements \cite{vn55b} and his proposal (together with
Birkhoff) of a special quantum logic \cite{bvn36}, neither of which are
employed
in the formulation of quantum theory presented here.  Next comes the concept of
a consistent history \cite{gr84}, followed by Omn\`es' proposal that quantum
reasoning must involve separate ``logics'' \cite{om}, and the Gell-Mann
and Hartle decoherence functional \cite{gmh}.  To this list the present
article adds one new idea, or at least a new word: the {\it framework} in which
a classical or quantum description is embedded, and relative to which its truth
must be assessed.

	A framework corresponds to a consistent family in the notation
of \cite{gr84}, or a ``logic'' in the notation of \cite{om}.  It is closely
related (but not identical) to the concept of an ``interpretation'' in
mathematical logic \cite{logic}.  The key to making sense out of the
description
of a quantum system, either at a single time or as it develops in time, is to
note that such a description cannot be made without adopting, at least
implicitly, some framework; that the truth of such a description is relative to
its framework; and that reasoning in the quantum domain requires the use of
{\it compatible} frameworks.  All of these concepts can be made quite precise,
and using them should help clear up (what I regard as) some misunderstandings,
and respond to various criticisms of the consistent history
program \cite{pz93,des95,dk95}.

	The present work is very much indebted to Omn\`es' ideas \cite{om}. A
crucial difference is that the notion of ``truth'' is made to depend explicitly
on the framework, following certain classical analogies, Sec.~\ref{s2a}, and
the example of formal logic.  By contrast, Omn\`es defined ``truth'' in terms
of ``facts'' which arise in a quasi-classical approximation, Sec.~\ref{world}.
The latter does not seem an entirely satisfactory approach for a fundamental
theory of nature, and certain criticisms have been made by
d'Espagnat \cite{des95}, and by Dowker and Kent \cite{dk95}.  Omn\`es
himself \cite{ompc} agrees that his approach has some problems.

	The conceptual structure presented in this paper appears adequate to
support the position of Englert et al. \cite{ess92} (the arguments are not
presented here), and to find the mistakes (from the perspective of
standard quantum mechanics) in at least some derivations of Bell's inequality;
see \cite{gr94}.  There are other derivations of Bell's inequality which rest
upon counterfactual reasoning: what {\it would} have happened {\it if}
something had been different; in addition, many quantum paradoxes involve a
counterfactual element.  Analyzing these will require an extension of the ideas
discussed here; see Sec.~\ref{s6b}.

	While its main goal is the clarification of conceptual issues through
the introduction of suitably precise rules of reasoning in the quantum domain,
this paper also contains, in Sec.~IV, some new results on quantum histories and
consistency conditions.  The representation of histories through the use of
projectors on tensor products of copies of the Hilbert space seems to be a new
idea, and the notion of a ``generalized'' history based upon this
representation includes, but also goes well beyond, previous proposals by the
author \cite{gr84,gr93}, Omn\`es \cite{om}, and Gell-Mann and
Hartle \cite{gmh}.  The ``consistency functional'' introduced in the same
context generalizes the ``decoherence functional'' of Gell-Mann and
Hartle \cite{gmh}, and results in a formulation of fundamental quantum theory
which is (transparently) invariant under reversing the direction of time.

	Philosophical issues are not the main topic of this paper.  However,
its central argument provides a number of details supporting a proposal, found
in an earlier publication \cite{gr93b}, for interpreting quantum mechanics in a
``realistic'' manner.

	Many of the crucial features which distinguish quantum reasoning from
its classical counterpart already arise when one considers a single mechanical
system at a single instant of time.  Hence Secs.~II and III, in which the
quantum description of a system at one time is developed with the help of
classical analogies, form the heart of this paper.  In particular, discussing
both classical and quantum systems from the same point of view helps avoid the
error of supposing that the latter somehow involve a subjective element not
present in the former, while bringing out the genuine differences between the
reasoning processes appropriate in the two situations.

	Describing how a quantum system evolves in time turns out to be,
formally at least, very similar to describing its properties at a single time
if one employs the technical tools in Sec.~IV, which allow a history to be
represented by a single projector on a tensor product space.  The notion of
{\it consistency} has, to be sure, no counterpart in a system at a single time,
and discussing it in a precise and general way makes Sec.~IV more complicated
than the other parts of the paper.  The reader who is unfamiliar with the use
of consistent histories in quantum mechanics should look elsewhere
 \cite{gr84,om,gr93} for a discussion of the physical motivation behind the
consistency requirement, and simple examples of its use.

	The applications in Sec.~V can be understood without the technical
machinery of Sec.~IV if the reader is willing to accept various results on
faith; the proofs are, in any case, not included in this paper.  These examples
are not new (except, perhaps, for the discussion of two-slit diffraction), but
have been chosen to illustrate the formulation of quantum reasoning introduced
earlier in the paper.  Sec.~VI contains a summary of this formulation, together
with a list of open questions.


		\section{Classical Reasoning}
\label{s2}
	\subsection{Frameworks and Descriptions}
\label{s2a}

	A scientific description of a physical system is, at best, an abstract,
symbolic representation of reality, or of what the scientist believes that
reality to be; the description is never reality itself.  Thus it necessarily
embodies certain elements of human choice.  Nonetheless, at least in the
``classical'' world of our everyday experience, it is not unreasonable to claim
that such a description is, or at least might be, a faithful or ``true''
representation of reality.  Understanding the process of description in the
classical realm will assist us in understanding how elements of choice can
enter quantum descriptions without necessarily making them any less
``objective'' than their classical counterparts.

	As a first example, consider representing a three-dimensional object,
such as a vase, by means of a two-dimensional drawing which shows a projection
of the object on a particular plane. It will be convenient to refer to the
choice of projection plane and the various conventions for representing salient
features of the object in the drawing as a {\it framework} $\F$, and the
drawing itself as a {\it statement}, $f$, with the pair $(\F,f)$ constituting a
{\it description} of the object.  The following are obvious properties of such
descriptions:
\begin{list}{}{}
\item[1.] A framework is chosen by the person making the description,
and without
such a choice no description is possible.
\item[2.] Choosing a framework has, by itself, no influence on the object being
described; on the other hand it constrains what can correctly be said about the
object. (Features which are visible in one projection may well be invisible in
another.)
\item[3.] A framework by itself is neither true nor false.  A statement can be
true (i.e., a correct representation of the object) or false, but to determine
which it is, one must know the corresponding framework.  That is, truth or
falsity is relative to the framework.
\item[4.] The correctness of a description can be tested experimentally.  (For
example, the distance between two points on the drawing can be checked by
measuring the distance between the corresponding features on the object,
provided one knows the projection plane and the scale of the drawing.)
\item[5.] An arbitrary collection of descriptions, each of which applies
individually to the object, can always be thought of as parts of a collective
or composite description of the object.
\end{list}

	Apart from the comments in parentheses, {\it every item in this list,
with the sole exception of 5, also applies to a quantum description\/}! That
is, quantum descriptions are no less ``objective'' or ``realistic'' than their
classical counterparts. However, they are different (reflecting, one might
suppose, the peculiar nature of quantum reality) in that it is not always
possible to combine various descriptions which might separately apply to a
particular object into a single composite description. This feature has no
classical analog (that we know of), and hence understanding what it means is
essential for consistent quantum reasoning.  In particular, one needs an
appropriate mathematical formalism, discussed in Secs.~III and IV below, giving
the precise rules for combining (or not combining) quantum descriptions, and
reasoning about them, together with examples, of which there are some in
Sec.~\ref{s5}, which illustrate the rules.

	Each of the five points in the preceding list can be illustrated using
as an example the intrinsic angular momentum or spin, measured in units of
$\hbar$, of a spin 1/2 quantum particle.  One possible description, which
employs the $x$ component of the spin, has the form:
\begin{equation}
	(S_x,1/2),
\label{e2.1}
\end{equation}
where $S_x$ is the framework and $1/2$ is the statement. Together they
form the  description usually written as $S_x=1/2$.  Another
equally good description is:
\begin{equation}
	(S_x,-1/2),
\label{e2.2}
\end{equation}
or $S_x=-1/2$.  Because (\ref{e2.1}) and (\ref{e2.2}) employ the same
framework, the same (framework dependent!) notion of ``truth'' applies to both.
In fact, if (\ref{e2.1}) is true, (\ref{e2.2}) is false, and {\it vice versa}.
Choosing the framework $S_x$, i.e., choosing to talk about $S_x$, does not
make $S_x=1/2$ true or false, in fact, it has no effect whatsoever on the spin
of the particle.  However, such an assertion is subject to experimental
verification (or falsification) ``in principle'', that is to say, in terms of
idealized experiments which, while they may not be practical, at least do not
violate the principles of quantum theory itself.  Thus we might imagine that
the spin 1/2 particle is a neutron traveling slowly towards a Stern-Gerlach
apparatus whose magnetic field gradient is in the $x$ direction, equipped with
a pair of counters to determine the channel in which the neutron emerges.  If
the neutron later emerges in the $+1/2$ channel, that will verify the
correctness of the description $S_x=1/2$ at the present time, before the
neutron enters the field gradient.  Even the ``classical'' problem of checking
a two-dimensional projection of a three-dimensional object by making
measurements on the object requires a certain theoretical analysis: relating
the distance between two points on the drawing to the measured distance between
the corresponding features on the object requires knowing both the projection
plane and the scale factor of the drawing, and using some trigonometry.
Similarly, experimental verification of a quantum description requires an
appropriate theory of measurement, as discussed in Sec.~\ref{measure}.

	We now come to item 5 in the list.  The description
\begin{equation}
	(S_z,-1/2),
\label{e2.3}
\end{equation}
or $S_z=-1/2$, which uses the $z$ component of the spin, is on a par with
(\ref{e2.1}) or (\ref{e2.2}).  It simply represents a ``projection'' of the
particle's angular momentum on a different axis.  Were we dealing with a
classical spinning object there would be not the slightest problem in combining
its various components of angular momentum into a complete description
corresponding to a three-dimensional vector.  But in the quantum case, the
frameworks $S_x$ and $S_z$ are {\it incompatible}, a term which will later
receive a precise definition, and as a consequence one of the basic rules of
quantum descriptions asserts that {\it there is no way to combine $S_x=1/2$ and
$S_z=-1/2$ into a single description}.  Note that this is {\it not} the
assertion that if $S_x=1/2$ is ``true'', then $S_z=-1/2$ is ``false''.  Truth
and falsity are concepts relative to a particular framework, and in the case of
descriptions using $S_x$ and $S_z$, there is no single framework in which a
common notion of truth can be applied to both of them.

	The issue of compatibility of frameworks arises because standard
quantum mechanics, unlike classical mechanics, employs a Hilbert space, rather
than a classical phase space, for descriptions.  Incompatibility reflects the
fact that certain operators do not commute with each other, a peculiarly
quantum phenomenon.  In terms of a spin 1/2 particle, the Hilbert space
structure manifests itself in the following way.  Descriptions (\ref{e2.1}),
(\ref{e2.2}), and (\ref{e2.3}) correspond, as is well known, to particular
one-dimensional subspaces or rays of the two-dimensional Hilbert space $\H$
which represents the spin of the particle; in fact, they are the subspaces
spanned by the corresponding eigenvectors of the operator $S_x$ or $S_z$.  But
as there are no simultaneous eigenvectors of $S_x$ and $S_z$, there are no
subspaces of $\H$ which might be thought of as corresponding to $S_x=1/2$ {\it
and} $S_z=-1/2$, that is, to the conjunction of these two descriptions.
Another way of thinking about the matter is to note that in classical
mechanics, specifying the $z$ component of angular momentum provides additional
information beyond that obtained through specifying the $x$ component, whereas
for a spin 1/2 particle one cannot specify any information in addition to
$S_x=1/2$, which is already a pure state.

	It is helpful to think of a framework, classical or quantum, as
defining a ``topic of conversation'' and, just as in ordinary conversation, in
scientific discourse the framework is often chosen implicitly. The phrase
``oranges cost 25 cents each at the supermarket'' both defines a topic of
conversation and simultaneously asserts something about the nature of the
world, and the same is true of ``$S_x=1/2$''. While it is hard to imagine any
efficient scheme of scientific communication which did not define frameworks
implicitly most of the time, it is sometimes helpful, especially in the quantum
case, to reflect upon just what framework is being used at some point in a
discussion, as this can help avoid the nonsensical statements and general
confusion which arises when an invalid change of framework occurs in the middle
of an argument.

	\subsection{Classical Phase Space}
\label{s2b}

	The usual framework for discussing a classical mechanical system of $N$
particles at a single instant of time is provided by its $6N$-dimensional phase
space $D$, in which a single point $x$ represents the state of the system at
the time in question.  A statement $p$, such as: ``the total energy is between
10 and 11 ergs'', can then be thought of as associated with the subset
\begin{equation}
	P=\varphi(p)
\end{equation}
of points in $D$ where $p$ is true, that is, the set contained between two
surfaces of constant energy, one at 10 ergs and the other at 11 ergs.  The
negation of $p$, denoted $\sim\! p$, corresponds to $D\smallsetminus P$, the
complement of $P$ in $D$, the region in the phase space for which the energy is
less than 10 ergs or greater than 11 ergs.  Another statement $q$:
``there are exactly 35 particles in the volume element $V_1$'', where $V_1$
denotes some definite region in three-dimensional space, corresponds to the
subset $Q=\varphi(q)$ of $D$.  The conjunction of $p$ and $q$, ``energy between
10 and 11 ergs {\it and} 35 particles in $V_1$'',written $p\wedge q$,
corresponds to the intersection $P\cap Q$ of the sets $P$ and $Q$.  	Thus
one sees that various ``elementary'' statements which describe properties of
the system can be mapped onto subsets of $D$, and other ``composite''
statements
formed from the elementary statements by use of the {\it logical operations}
``not'', ``and'', ``or'', denoted respectively by $\sim,\wedge,\vee$, are then
also mapped to subsets of $D$ using complements, intersections, and unions of
the sets corresponding to the elementary statements.  Note that under the
operations of complement, union, and intersection, the subsets of $D$ form a
Boolean algebra $\B$.

	It will be convenient to think of the {\it framework} $\cal F$ as
consisting of elementary and composite statements, together with the set $D$,
the Boolean algebra $\B$ of its subsets, and the function $\varphi$ which maps
the statements onto elements of $\B$.  A description $(\F,f)$, where $f$ is one
of the statements, corresponds to the assertion that the state of the
mechanical
system is one of the points in the subset $F=\varphi(f)$, and is true or false
depending upon whether the point $x$ corresponding to the actual state of the
system at the time in question is or is not in $F$.  Note that many different
statements may correspond to the same $F$; for example, $p\vee\!\sim\!p$ for
any statement $p$ is mapped onto the entire phase space $D$. When it is not
important to distinguish the different possibilities corresponding to some $F$,
one may use a description $(\F,F)$, with $F$  a surrogate for any
statement which is mapped onto it by $\varphi$.

	A description formed in this way is what in first-order predicate logic
 \cite{logic} is called an {\it interpretation}.  For our purposes it suffices
to summarize the rules of logical reasoning, {\it modus ponens} and the like,
as applied to descriptions of our classical mechanical system, in the following
way.  The logical process of inference from a set of {\it assumptions} to valid
{\it conclusions} consists in taking the intersection, let us call it $A$, of
all the subsets of $D$ corresponding to all of the assumptions---if one
prefers, $A$ is the image under $\varphi$ of the single statement consisting of
the conjunction (``and'') of all the assumptions.  Then any conclusion which
corresponds to a subset $C$ of $D$ which contains $A$ is a valid conclusion
from these assumptions.  That is to say, logical reasoning in this context is
the process of checking that if there is any $x$ in $D$ for which all the
assumptions are true, then for this $x$ the conclusion will also be true.  Note
that the process of inference can begin with no assumptions at all, in which
case one sets $A$ equal to $D$, and a valid conclusion is a tautology, such as
$p\vee\!\sim\!p$.



		\section{Quantum Systems at One Time}
\label{s3}

	\subsection{Quantum Frameworks}
\label{s3a}

	The counterpart for a quantum system of the classical phase space $D$
is a Hilbert space $\cal H$, and an elementary statement $p$ which ascribes to
the quantum system a property at a particular time is associated with a closed
subspace $\cal P$ of $\cal H$ or, equivalently, the {\it projector} (orthogonal
projection operator) $P$ onto this subspace.  For example, if $p$ is the
assertion that ``the total energy is between 10 and 11 ergs'', the subspace
$\cal P$ is spanned by the eigenvectors of $\cal H$ with eigenvalues (assumed,
for simplicity, to be discrete) which lie between 10 and 11 ergs.  The negation
$\sim\!p$, which asserts that the energy lies outside this interval,
corresponds to the orthogonal complement ${\cal P}^\bot$ of $\cal P$ (not to be
confused with ${\cal H}\smallsetminus{\cal P}$), with projector $I-P$, where
$I$ is the identity operator on $\cal H$.

	A {\it framework} $\cal F$ in the quantum case is generated by a finite
collection of elementary statements associated with projectors onto closed
subspaces of $\cal H$ {\it provided these projectors commute with each other}.
The projector associated with a statement $p$ will be denoted by $\varphi(p)$.
The additional statements belonging to $\cal F$ are produced from the
elementary statements using logical operations, as discussed in Sec.~II, and
are mapped by $\varphi$ onto projectors consistent with the rules
\begin{equation}
 	\varphi(\sim\!p) = I- \varphi(p),\ \
	\varphi(p\wedge q)=\varphi(p)\varphi(q).
\label{e3.1}
\end{equation}
(These rules suffice, because every logical
operation can be built up using ``not'' and ``and''.)  The smallest family
$\cal B$ of commuting projectors closed under complements and products, and
containing the projectors associated with the statements of $\cal F$, is a
Boolean algebra in which the operations of $\cap$ and $\cup$, thought of as
acting on pairs of projectors, are defined by
\begin{equation}
	P\cap Q = PQ, \ \
	P\cup Q = P + Q -PQ,
\label{e3.2}
\end{equation}
and whose least and largest elements are $\emptyset$ (the zero operator) and
$I$, respectively.  The framework $\cal F$ can then be defined as the
collection of statements generated from a set of elementary statements by
logical operations, along with the Hilbert space $\cal H$ and the mapping
$\varphi$ which
carries statements onto the Boolean algebra $\cal B$ of commuting projectors
in the manner indicated above.

	A quantum {\it description} $(\F,f)$ consists of a framework $\F$ and a
statement $f$ belonging to the collection of statements associated with $\F$.
Sometimes it is convenient to use a description $(\F,F)$, where the projector
$F$ is an element of the Boolean algebra $\B$ associated with $\F$, and serves
as a surrogate for any $f$ mapped to this $F$ by $\varphi$.  As long as $\F$ is
held fixed, there is a close analogy with classical descriptions based upon a
phase space as discussed in Sec.~\ref{s2b} above.  Since all the projectors in
the Boolean algebra $\B$ associated with $\F$ commute with each other, it is
possible to choose a representation in which they are simultaneously diagonal,
which means that each diagonal element $\lg j|P|j\rg$ of a projector
$P=\varphi(p)$ in $\B$ is either 0 or 1. If one thinks of the set of labels
$\{j\}$ of the diagonal elements as constituting a set $D$, then the subset of
$D$ where $\lg j|P|j\rg=1$ is analogous to the set of points in the classical
phase space $D$ where the statement $p$ is true.  This analogy works very well
as long as $\F$ is held fixed, and permits one to construct the rules for
quantum reasoning in close analogy with their classical counterparts.

	In particular, one thinks of a quantum description $(\F,p)$ as {\it
true} provided the quantum system at the time in question is ``in'' the
subspace onto which $P=\varphi(p)$ projects.  To be more precise, if we assume
that any projector $B$ belonging to the Boolean algebra $\B$ associated with
$\F$ represents a statement about, or property of the quantum system which is
true at the time in question, then this implies the truth of any statement $s$
with the property that
\begin{equation}
	BS=B,
\label{e3.3}
\end{equation}
where $S=\varphi(s)$ is the projector corresponding to $s$, that is, provided
the subspace onto which $B$ projects is contained in the subspace onto which
$S$ projects.  Just as in the case of the classical descriptions discussed in
Sec.~\ref{s2}, the truth of a quantum description must always be thought of as
relative to a framework, which may have been defined implicitly.  Hence it is
best to think of ``$(\F,f)$ is true'' as meaning: ``Given $\F$, then $f$ is
true.''

	Quantum reasoning within the context provided by a single framework
$\F$ proceeds in the following way.  The assumptions of the argument, $ a_1,
a_2,\ldots a_l$, which must all be statements belonging to $\F$, are
mapped by
$\varphi$ onto a set of projectors $A_1, A_2, \ldots A_l$ belonging to $\B$,
whose product is:
\begin{equation}
	A=A_1 A_2 \cdots A_l.
\end{equation}
Then a statement $c$ in $\F$ is a valid conclusion provided $AC=A$, where
$C=\varphi(c)$.

	\subsection{Compatible and Incompatible Frameworks}
\label{s3b}

	The most important differences between classical and quantum
descriptions emerge when one considers several different frameworks.  In the
classical case, as long as the frameworks refer to the same system, there is no
problem in combining the corresponding descriptions.  But in the quantum case
this is no longer true, and it is necessary to pay attention to the rules which
state when descriptions can and cannot be combined.

	A finite collection of frameworks $\{{\cal F}_i\}, i=1,2,\ldots l$,
will be said to be (mutually) {\it compatible} if each framework employs the
same Hilbert space ${\cal H}$, and if all the projectors associated with the
different Boolean algebras ${\cal B}_i$ {\it commute with one another}.  In
addition, the statements belonging to the different frameworks should be mapped
onto the projectors in a consistent way, so that an elementary statement which
occurs in more than one framework is mapped to the same projector.  (This last
point is a matter of notational consistency which causes little difficulty in
practice.)  Given a compatible collection $\{{\cal F}_i\}$ there is a smallest
framework $\cal F$ which contains them all: its statements are generated from
the union of the sets of elementary statements for the individual ${\cal F}_i$,
and its Boolean algebra $\cal B$ of projectors is the smallest one containing
all the projectors of all the Boolean algebras ${\cal B}_i$ associated with the
different frameworks in the collection.  We shall say that the compatible
collection $\{{\cal F}_i\}$ {\it generates} this smallest framework $\cal F$.

	Two or more frameworks which are not compatible are called {\it
incompatible}.  The distinctive problems associated with quantum reasoning
arise from the existence of frameworks which use the same Hilbert space, and
can thus (potentially) refer to the same physical system, but which are
mutually incompatible because the projectors associated with one framework do
not commute with those of another.  There is nothing quite like this in
classical mechanics, since the classical counterparts of projectors always
commute with one another.

	We adopt the following as a fundamental rule. {\it A meaningful quantum
description must consist of a single framework and one of its statements,
$(\F,f)$, or else a compatible collection of frameworks $\{\F_i\}$ and
associated statements $\{f_i\}$, which together form a collective description
$\{(\F_i,f_i)\}$}.
	A collective description can always be replaced by a single
{\it master description} $(\D,d)$, where $\D$ is the framework generated by the
collection $\{\F_i\}$, and $d$ is any statement such that
\begin{equation}
	D=\varphi(d) = F_1 F_2 \cdots F_l
\label{e3.D}
\end{equation}
is the product of the projectors $F_i=\varphi(f_i)$ corresponding to the
different statements.

	The rules for logical reasoning about quantum descriptions are similar
to the rules for reasoning about classical descriptions, but with the
additional requirement that all the frameworks must be compatible.  A logical
{\it argument} begins with a a set of {\it assumptions}, which is to say a set
of descriptions $\{(\A_i,a_i)\}, i=1,2,\ldots l$, associated with a {\it
compatible} family $\{\A_i\}$ of frameworks; one is assuming that these
descriptions are simultaneously true (within the framework generated by the
collection $\{\A_i\}$).  From these assumptions one can deduce a set of valid
{\it conclusions} $\{(\C_j,c_j)\}, j=1,2,\ldots m$, provided the union of the
collections $\{\A_i\}$ and $\{\C_j\}$ is a compatible collection of frameworks,
and
\begin{equation}
	\varphi(c_j)A=A,
\end{equation}
holds for every $j$, where
\begin{equation}
	A=A_1 A_2 \cdots A_l
\label{e3.A}
\end{equation}
is the product of the projectors $A_i=\varphi(a_i)$.  The set of assumptions
can always be replaced by the corresponding master description acting as a
single assumption, since the process of reasoning just described allows one to
deduce the original set of assumptions from this master description.  Avalid
argument can begin with the true assumption $(\F,I)$, where $\F$ is any
framework, and $I$ is the identity operator.

	It is extremely important that within a {\it single} argument, {\it
all} the frameworks, for both the assumptions and conclusions, be {\it
compatible.} In particular, the following sort of ``quasi-classical
reasoning'', many examples of which are to be found in the published
literature, is not valid: One starts with an assumption $(\A,a)$, and from it
deduces a conclusion $(\C,c)$, after checking that the frameworks $\A$ and $\C$
are compatible.  Next, $(\C,c)$ is used as the assumption in an argument whose
conclusion is $(\E,e)$, again after checking that the frameworks $\C$ and $\E$
are compatible.  Combining these two arguments one draws the conclusion that
``is $a$ is true, then $e$ must be true''.  But this reasoning process is only
correct if $\A$ and $\E$ are compatible frameworks; if they are incompatible,
it is invalid.  (Note that compatibility is not a transitive relationship: $\A$
can be compatible with $\C$ and $\C$ with $\E$ at the same time that $\A$ is
incompatible with $\E$.)
Failure to check compatibility can easily lead to inconsistent
quantum reasoning which, just like inconsistent classical reasoning, often
produces contradictions and paradoxes.  In order to avoid making mistakes of
the sort just described, it is helpful to think of a single quantum argument as
a process in which the assumptions, which themselves form a (collective)
description, are extended into a longer and longer collective description by
adding additional simple descriptions without ever erasing anything; in
particular, without forgetting the original assumptions. The compatibility rule
for collective quantum descriptions will then rule out any attempt to introduce
incompatible frameworks.

	On the other hand, two or more {\it separate} arguments may very well
involve different, and perhaps incompatible frameworks.  But then they cannot
be combined into a single argument.  As an example, it is possible to construct
{\it two} perfectly valid arguments, each based upon the {\it same} assumption
$(\A,a)$, one leading to the conclusion $(\C,c)$ and the other to the
conclusion $(\E,e)$, where the frameworks $\C$ and $\E$ are incompatible. If
one grants the truth of the assumption, then $c$ is true relative to framework
$\C$, and $e$ is true relative to framework $\E$. However, there is no
framework relative to which it can be said that both $c$ and $e$ are true
statements. Given a particular physical system for which the assumptions are
satisfied, it is possible in principle to check the validity of either $c$ or
$e$ by means of measurements, but not (at least in general) the validity of
both.  An example (involving histories) which illustrates this point will be
found in Sec.~\ref{between} below.

	If one wishes to describe a quantum system, it is necessary to choose a
framework, if only implicitly, and in all but the most trivial cases this
choice means that there are many incompatible frameworks whose statements
cannot be employed in the description.  While the choice of a framework does
not in any way ``influence'' the system being described, Sec.~\ref{s2a}, it
severely constrains what can sensibly be said about it.  Thus if one wants to
talk about the energy, this requires the use of certain projectors, depending
on how precisely one wants to specify the energy, and what ranges one is
interested in.  It is then not possible, as part of the same description, to
talk about something else represented by projectors which do not commute with
the set employed for the energy.

	The type of exclusion which arises from incompatible frameworks is
easily confused with, but is in fact quite different from the sort of exclusion
which arises all the time in classical mechanical descriptions, where in order
that some property $p$ be true, it is necessary that some other property $q$ be
false (because the corresponding subsets of the classical phase space do not
overlap, Sec.~\ref{s2b}).  This ``classical'' type of exclusion also arises in
quantum systems where, to take the example of a spin 1/2 particle,
``$S_x=1/2$'' and ``$S_x=-1/2$'' are mutually exclusive statements belonging to
the same framework: if one is true, the other is false.

	By contrast, if $p$ and $q$ are assertions about a particular quantum
system represented by projectors $P$ and $Q$ which do not commute, they cannot
be part of the same framework.  Hence in a framework $\P$ in which it makes
sense to talk about the truth and falsity of $p$, there is no way of discussing
whether $q$ is true or false, since within this framework $q$ makes no sense.
Conversely, in a framework $\Q$ in which $q$ makes sense, and could be true or
false, $p$ has no meaning.  Thus the truth of $p$ does not make $q$ false; it
does mean that because we have (perhaps implicitly) adopted a framework in
which it makes sense to talk about $p$, $q$ cannot be discussed.  Similarly,
the combination $p\wedge q$, or ``$p$ and $q$'', is not part of any framework,
and therefore cannot be true or false.  It is ``meaningless'' in the sense that
within quantum theory one cannot ascribe any meaning to it.  In mathematical
logic there are certain combinations of symbols, for example $p\wedge\vee q$,
which are ``nonsense'' because they are not ``well-formed formulas'', they are
not constructed according to the rules for combining symbols of the language to
form meaningful statements.  In the quantum case, where the rules are, of
course, different from those of classical logic, $p\wedge q$ (and also $p\vee
q$, etc.) has this nonsensical character when $PQ\neq QP$.

		\section{Quantum Systems at Many Times}
\label{s4}
	\subsection{Classical Analogy}
\label{s4a}

	A classical stochastic process, such as rolling a die several times in
succession, is a more useful analogy to quantum time dependence than is a
continuous trajectory in a classical phase space.  If a die is rolled three
times in succession, the possible outcomes are triples of numbers
$(d_1,d_2,d_3),\ 1\leq d_j\leq 6$, which together constitute the set
\begin{equation}
	\tilde D = D^3
\label{e4.1}
\end{equation}
of $216=6^3$ histories, the sample space of classical probability theory.  Here
$D$ is the set of possible outcomes of one toss.  A statement such as $p$:
``the sum of the first two tosses is four'', is associated with the subset
$\tilde P =\varphi(p)$ of $\tilde D$ consisting of those histories for which
$p$ is true.  The process of logical inference employing statements of this
kind is then formally identical to that discussed in Sec.~II.B.  For example,
from $p$ one can immediately infer $r$: ``the sum of all three tosses is five
or more'' by noting that $\tilde P$ is a subset of $\tilde R = \varphi(r)$, the
set of histories where $r$ is true.

	On the other hand, the statement $q$: ``the sum of all three tosses is
exactly seven'', is not a logical consequence of $p$, since $\tilde
Q=\varphi(q)$ is a subset of $p$.  However, if probabilities are assigned to
$\tilde D$, one can compute the {\it conditional probability}
\begin{equation}
	\Pr(q\v p)=W(\Qt\cap\Pt)/W(\Pt)
\label{e4.2}
\end{equation}
of $q$ given $p$, using {\it weights} $W(\At)$ for subsets $\At$ of $\Dt$.  In
the case of an unloaded die, $W(\At)$ is just the number of elements
(histories) in the set $\At$. On the other hand, making appropriate
replacements in (\ref{e4.2}), one sees that $\Pr(r\v p)=1$.  Indeed, as long as
every history has a finite weight, there is a close connection between $r$
being a {\it logical} consequence of $p$, and $\Pr(r\v p)=1$; the only
difference comes about in the case in which the statement $p$ corresponds to an
empty set with zero weight, for which the right side of (\ref{e4.2}) is
undefined.

	\subsection{Histories and Projectors}
\label{s4b}

	The simplest type of {\it history} for a quantum system with Hilbert
space $\H$ is constructed in the following way. Let a finite set of times $t'_1
< t'_2 < \cdots < t'_j$ be given, and at each $t'_j$ let $P'_j$ be a projector
onto a closed subspace of $\H$.  The history consists of the set of times and
the associated projectors, representing properties of the system which are true
at the corresponding times.  So that this history can be discussed within
the same framework as other histories defined at different sets of times, it is
convenient to suppose that there is a common set of times
\begin{equation}
	t_1 < t_2 < \cdots  < t_n
\label{e4.t}
\end{equation}
sufficiently large to include all those associated with the different histories
in the framework, and that a particular history is represented by a sequence of
projectors $P_1, P_2,\ldots P_n$, one for each of the times in (\ref{e4.t}). To
do this, $P_k$ is set equal to the identity $I$ if $t_k$ is not one of the
times for which the history was originally defined, and to $P'_j$ if
$t_k=t'_j$.  Since the ``property'' represented by $I$ is always true, it may
be ``added'' to the original history at additional times without modifying its
physical significance.

	Rather than representing a history with a sequence of projectors $P_1,
P_2,\ldots P_n$, it is technically more convenient to employ a {\it single}
projector
\begin{equation}
	\tilde P = P_1\otimes P_2\otimes \cdots \otimes P_n
\label{e4.pr}
\end{equation}
on the tensor product space
\begin{equation}
	\tilde{\cal H} = \cal H \otimes H\otimes \cdots \otimes H
\label{e4.hp}
\end{equation}
of $n$ copies of $\cal H$, which is analogous to $\Dt$ in the classical
case discussed earlier.  It is obvious that if each of the $P_j$ on the
right side of (\ref{e4.pr}) is a projector, so is $\tilde P$; on the
other hand,
not every projector on $\tilde{\cal H}$ has the form (\ref{e4.pr}).  It will be
convenient to extend the concept of a history to include {\it any} projector on
$\tilde{\cal H}$.  The resulting category of {\it generalized histories} is
large enough to include the proposals made by Omn\`es \cite{om}
(``non-Griffiths histories''), and by Gell-Mann and Hartle \cite{gmh}
(``history-dependent decompositions of the identity''), and much else besides.
However, the applications discussed in Sec.~V are all of the restricted form
(\ref{e4.pr}), hereafter referred to as {\it simple histories\/}.

	A {\it framework} $\cal F$ for the case of many times can be
constructed in the following way.  We suppose that there are a finite number of
elementary statements $p,q,\ldots$, and that to each of these there corresponds
a projector on $\tilde{\cal H}$, representing some (generalized) history.  As
in the case of a quantum system at one time, we require that all these
projectors commute with each other.  They are therefore members of a smallest
Boolean algebra $\tilde{\cal B}$ of commuting projectors closed under the
operations of products and complements, where $\tilde I -\tilde P$ is the
complement of $\tilde P$, and
\begin{equation}
	\tilde I = I \otimes I\otimes \cdots \otimes I
\label{e4.4}
\end{equation}
is the identity on $\tilde{\cal H}$.  Additional statements formed from the
elementary statements by logical operations will also be mapped onto elements
of $\tilde{\cal B}$ following the rules in (\ref{e3.1}).  Thus the
structure is formally the same as that in Sec.~III.  However, for a framework
involving histories to be acceptable it must also satisfy consistency
conditions, which is the next topic.

	\subsection{Consistency and Weights}
\label{s4c}

	Various consistency conditions have been proposed by various authors;
what follows is closest in spirit to \cite{gr93}, while very much indebted to
the ideas in \cite{gmh}.  In the following discussion it will be convenient to
assume that the Hilbert space $\H$ is {\it finite dimensional}, so that there
are no questions about convergence of sums.  This will not bother low-brow
physicists who are quite content to consider a quantum system in a box of
finite volume with an upper bound (as large as one wishes) on the energy.
Extending the formalism in a mathematically precise way to an
infinite-dimensional $\cal H$ remains an open problem.

	Let $T(t',t)$ be the unitary time transformation which represents
Schr\"odinger time evolution from time $t$ to time $t'$. In the case of a
time-independent Hamiltonian $H$ it is given by:
\begin{equation}
	T(t',t)=\exp[-i(t'-t)H/\hbar].
\label{e4.5}
\end{equation}
The discussion which follows is not restricted to time-independent
Hamiltonians, but we do require that $T(t',t)$ be unitary and satisfy the
conditions
\begin{equation}
	\begin{array}{rcl}
	T(t,t)& =& I,\\
	T(t,t')T(t',t'')&=& T(t,t''),
	\end{array}
\label{e4.6}
\end{equation}
which imply that $T(t,t')$ is the inverse of $T(t',t)$.

	Given $T(t',t)$, one can define a mapping from an arbitrary operator
$\tilde A$ on $\H^n$ to an operator $K(\tilde A)$ on $\H$ by means of the
formula:
\begin{equation}
	\begin{array}{rl}
\lg i|K(\tilde A)|j\rg & =\sum\limits_{k_2}\sum\limits_{k_3}\ldots
\sum\limits_{k_n}\sum\limits_{l_1}\sum\limits_{l_2}\ldots\sum\limits_{l_{n-1}}
	 \lg i k_2 k_3 \ldots k_n|\tilde A|l_1 l_2\ldots l_{n-1} j\rg \\
	&\times \lg l_1|T(t_1,t_2)|k_2\rg \lg l_2|(T(t_2,t_3)|k_3\rg \cdots
\lg l_{n-1} |T(t_{n-1},t_n)| k_n\rg,
	\end{array}
\label{e4.7}
\end{equation}
where the matrix elements refer to some orthonormal basis $\{|j\rg\}$ of $\H$
and the corresponding tensor product basis of $\H^n$. If, in particular,
$\tilde A$ is a projector $\tilde P$ of the form (\ref{e4.pr}), then $K(\tilde
P)$ takes the form:
\begin{equation}
	K(\tilde P) = P_1 T(t_1,t_2) P_2 T(t_2, t_3) \cdots T(t_{n-1},t_n) P_n.
\label{e4.8}
\end{equation}
Using $K$, we define the bilinear {\it consistency functional $C$} on pairs of
operators $\tilde A$, $\tilde B$ on $\tilde\H$ by means of the formula:
\begin{equation}
	C(\tilde A, \tilde B) = C^*(\tilde B, \tilde A)=
	{\rm Tr}[K^\dagger(\tilde A) K(\tilde B)].
\label{e4.cf}
\end{equation}
Note that $C(\tilde A, \tilde A)$ is non-negative. If
both $\tilde A$ and $\tilde B$ are projectors of the form (\ref{e4.pr}), the
consistency functional is the same as the Gell-Mann and Hartle decoherence
functional if the density matrix in the latter is replaced by $I$.

	Because $\HT$ is finite-dimensional, the Boolean algebra $\BT$
associated with $\F$ contains a finite number of projectors, and consequently
there are a set of non-zero {\it minimal elements} $\{\Mt^{(\alpha)}\}$,
$\alpha=1,2,\ldots$ which form a decomposition of the identity,
\begin{equation}
	\tilde I = \sum_\alpha \tilde M^{(\alpha)},\ \
\tilde M^{(\alpha)} \tilde M^{(\beta)}=
\delta_{\alpha\beta} \tilde M^{(\alpha)},
\label{e4.di}
\end{equation}
in terms of which any projector $\Pt$ in $\BT$ can be written in the form
\begin{equation}
	\Pt = \sum_\alpha m_\alpha \tilde M^{(\alpha)},
\label{e4.}
\end{equation}
with each $m_\alpha$ equal to $0$ or $1$, depending on $\Pt$.
An acceptable framework $\cal F$ must satisfy the following {\it consistency
condition} in terms of these minimal elements:
\begin{equation}
	C(\tilde M^{(\alpha)},\tilde M^{(\beta)}) = 0
	\ {\rm whenever}\ \alpha \neq \beta.
\label{e4.c}
\end{equation}
If this is satisfied, a
non-negative {\it weight} $W(\tilde P)$ can be defined for every element of
$\BT$ through the formula:
\begin{equation}
	W(\tilde P) = C(\tilde P,\tilde P),
\label{e4.w}
\end{equation}
and these weights can be used to generate conditional probabilities in analogy
with the classical case, (\ref{e4.2}), and as discussed in \cite{gr84}
and \cite{gr93}.  In particular, with $\Pt=\varphi(p)$ and $\Qt=\varphi(q)$ the
projectors associated with $p$ and $q$, the counterpart of (\ref{e4.2}) is the
formula
\begin{equation}
	\Pr(q\v p)=W(\Pt\Qt)/W(\Pt),
\label{e4.cp}
\end{equation}
with $W$ defined by (\ref{e4.w}).

	If only a single time is involved, then it is easy to show that the
consistency condition (\ref{e4.c}) is automatically satisfied.  It is also
automatically satisfied for all families of {\it simple} histories involving
only two times, $n=2$.  However, in the case of generalized histories involving
two or more times, and simple histories involving three or more times, the
consistency condition is a non-trivial restriction on acceptable families.

	The consistency functional (\ref{e4.cf}), unlike the decoherence
functional of Gell-Mann and Hartle, does not involve a density matrix.  This
means that the former, unlike the latter, does not single out a direction of
time. There seems to be no good reason why a density matrix should appear in a
fundamental formulation of quantum mechanics, especially since its classical
analog is properly part of classical statistical mechanics rather than
classical mechanics as such. To be sure, a density matrix may also arise in
quantum mechanics, unlike classical mechanics, as a technical device for
describing the state of a subsystem of a total system when the latter is in a
pure state.  But it is not needed for describing a single closed system, which
is what we are considering here. (Also see the remarks in Sec.~\ref{world}.)

	\subsection{Frameworks, Compatibility, and Logical Inference}
\label{s4d}

	The discussion above can be summarized by saying that a framework $\cal
F$ for a quantum system at several times, usually called a {\it consistent
family of histories}, consists of a collection of statements generated from a
set of elementary statements by logical operations, together with a mapping
$\varphi$, conforming in an appropriate way to the logical operations, of these
statements onto a Boolean algebra of commuting projectors on $\tilde{\cal H}$,
with the additional requirement that the minimal elements of this algebra
satisfy the consistency condition (\ref{e4.c}).

	A {\it description} is then a pair $(\F,f)$, consisting of the
framework or consistent family $\F$ and one of its statements $f$.  We shall,
with a certain lack of precision, call both the description $(\F,f)$ and the
corresponding statement $f$ a quantum {\it history}.  This ought not to cause
confusion if one remembers that statements by themselves do not have a meaning
(and, in particular, cannot be true or false) unless they are embedded in or
associated with some framework, which can either be specified explicitly by
giving the pair $(\F,f)$, or implicitly.  In particular, if $\Ft$ is the
projector corresponding to $f$, then the smallest Boolean algebra containing
$\Ft$ defines, in the absence of any other information, an implicit framework
for $f$, and this algebra will be part of the Boolean algebra of any consistent
family which has $f$ as one of its statements. (Note that for a particular
projector $\Ft$, even this smallest Boolean algebra may not satisfy the
consistency condition (\ref{e4.c}); in such a case $f$ is an inconsistent
(meaningless)  history, which cannot be part of any consistent family.)

	The intuitive significance of a history $(\F,f)$ is very similar to
that of a description at a single time, as discussed in Sec.~\ref{s3}.  In
particular, in the case of a simple history corresponding to (\ref{e4.pr}), one
thinks of the quantum system as actually possessing the property (represented
by) $P_j$ at each time $t_j$, if this history is the one which actually occurs.
Quantum mechanics, as a stochastic theory, cannot (in general) guarantee that
such a history takes place; instead, it assigns it a probability, based on some
assumption, such as the occurrence of the initial state, using the weights
generated by the consistency functional.  An empirical check on these
probabilities is possible ``in principle'', i.e., by idealized measurements
which do not violate the principles of quantum theory; see the extensive
discussion in \cite{gr84}.

	For a finite collection $\{{\cal F}_i\}$ of consistent families to be
(mutually) {\it compatible}, they must, to begin with, refer to a common tensor
product $\tilde{\cal H}$, (\ref{e4.hp}), or else it must be possible to achieve
this by defining $\tilde{\cal H}$ using a larger collection of times than those
found in the individual families, and then extending the latter by adding
identity operators to the histories at the times not previously considered.
Next comes a requirement of notational consistency: statements belonging to the
different frameworks must be mapped onto the projectors in a consistent way, so
that an elementary statement which occurs in more than one framework is mapped
to the same projector. There are then two additional and specifically quantum
conditions.  The first is the same as in Sec.~III.A: projectors corresponding
to the different Boolean algebras $\BT_i$ of the different families must
commute with each other.  The second is the consistency condition (\ref{e4.c})
imposed on the minimal elements of the Boolean algebra $\BT$ associated with
the smallest consistent family $\cal F$ containing all the ${\cal F}_i$, the
family generated by the collection $\{{\cal F}_i\}$.

	Given these definitions, one has the same fundamental rule as in
Sec.~III.B; worded in terms of histories, it reads: {\it A meaningful quantum
history must consist of a single consistent family together with one of its
histories (statements), $(\F,f)$, or else a compatible collection of consistent
families $\{\F_i\}$ and associated histories $\{f_i\}$, which together form a
collective history (or description) $\{(\F_i,f_i)\}$}. A collective history can
always be replaced by a single {\it master history} $(\D,d)$, where $\D$ is the
consistent family generated by the collection $\{\F_i\}$, and $d$ is any
history corresponding to the projector $\Dt$ which is the product of the
projectors $\Ft_i$ corresponding to the histories $f_i$, as in (\ref{e3.D}).

	The rules for logical reasoning in the case of histories are similar to
those for a quantum system at one time, Sec.~\ref{s3b}, except that they are
based upon conditional probabilities constructed from ratios of weights, using
(\ref{e4.cp}).  An {\it argument} begins with a a set of {\it assumptions},
which is to say a set of histories or descriptions $\{(\A_i,a_i)\},
i=1,2,\ldots l$, associated with a {\it compatible} set $\{\A_i\}$ of
consistent families; one is assuming that these histories are simultaneously
true (within the consistent family generated by the collection $\{\A_i\}$).
{}From these one can deduce a set of valid {\it conclusions} $\{(\C_j,c_j)\},
j=1,2,\ldots m$, provided the union of the collections $\{\A_i\}$ and
$\{\C_j\}$ is a compatible collection of consistent families, and provided
\begin{equation}
	\Pr(c_j\v \At)=W(\Ct_j \At)/W(\At)= 1
\label{e4.ir}
\end{equation}
holds for every $j$, where $\Ct_j=\varphi(c_j)$, and
\begin{equation}
	\At=\At_1 \At_2 \cdots \At_l,
\label{e4.A}
\end{equation}
with $\At_i$ the projector associated with the history $a_i$.  In order for the
inference to be valid, we require that $W(\At)$ be positive, so the right side
of (\ref{e4.ir}) is defined.  The case $W(\At)=0$ corresponds to a probability
of zero that all of the assumptions in the argument are simultaneously
satisfied, and is thus similar to having contradictory hypotheses in ordinary
logic.  Excluding the possibility of making any inferences from probability
zero cases, as suggested here, seems intrinsically no worse than the solution
in ordinary logic, in which any statement whatsoever can be inferred from a
contradiction.

	These rules coincide with those given in Sec.~\ref{s3b} in the
particular case where $n=1$, assuming $\H$ is finite dimensional,
because $W(P)$
for a projector $P$ is then just the dimension of the space onto which $P$
projects. The case of contradictory assumptions, $A=0$, again forms an
exception, for the reasons discussed above.

	The remarks at the end of Sec.~\ref{s3b} about the necessity of
choosing a particular framework in order to describe a quantum system also
apply to a quantum history.  It is necessary to choose some consistent family,
if only implicitly, in order to describe the time development of a quantum
system, and while this choice has no influence upon the system itself, it
constrains what can sensibly be said about it.  The consistent family is not
itself either true or false, but a history can be true or false, i.e., occur or
not occur, as one of the possibilities within a consistent family.  As long as
one considers a single family, a particular history may well exclude another
history in the sense that the probability is zero that both occur.  But this
sort of exclusion is very different from that which arises when two histories
$f$ and $g$ are incompatible in the sense that there is no consistent family
which includes both of them.  In such a case, the occurrence of history $f$
does {\it not} imply that $g$ does not occur; what it means is that in order to
even talk about the occurrence of $f$, we must employ a framework or
consistent family in which it makes no sense to say whether or not $g$ occurs,
since $g$ is not one of the histories in this family, and could not be added
to this family without making it inconsistent.  For the same reason, statements
such as ``$f$ and $g$'', or ``$f$ or $g$'', make no sense when these histories
cannot both belong to a single consistent family.  Such combinations are like
improperly-formed formulas in mathematical logic, sequences of symbols which
cannot be interpreted because they are not constructed according to the rules
appropriate to the particular language under discussion.



		\section{Applications}
\label{s5}

	\subsection{Measurements}
\label{measure}

	As a simple example of a measurement, consider the process of
determining $S_x$, the $x$ component of spin, for a spin 1/2 particle using a
Stern-Gerlach apparatus equipped with detectors to determine in which
channel the
particle emerges from the region where there is a magnetic field gradient.  The
essential features of the unitary time transformation $T(t_2,t_1)$ from a time
$t_1$ before the measurement takes place to a time $t_2$ after it is completed
are, in an obvious notation:
\begin{equation}
	|\alpha,X\rg \ra |\alpha',X^+\rg, \ \
	|\beta,X\rg  \ra |\beta',X^-\rg,
\label{e5.m1}
\end{equation}
where $|\alpha\rg$ corresponds to the spin state $S_x=1/2$ for the particle, in
units of $\hbar$, $|\beta\rg$ to $S_x=-1/2$, $|X\rg$ to the ``ready'' state of
the apparatus before the particle has arrived, and $|X^+\rg$ and $|X^-\rg$ are
apparatus states which result if the particle emerges from the field gradient
in the $S_x=1/2$ or $S_x=-1/2$ channel, respectively.  One should think of
$|X^+\rg$ and $|X^-\rg$ as macroscopically distinct states corresponding to,
say, two distinct positions of a pointer which indicate the outcome of the
measurement.  The spin states $|\alpha'\rg$ and $|\beta'\rg$ at time $t_2$ are
arbitrary; they are not relevant for the measuring process we are interested
in.  Throughout the following discussion we employ a symbol outside a Dirac ket
to indicate the corresponding projector; for example:
\begin{equation}
	\alpha = |\alpha\rg\lg\alpha|,\ \
	X^+=| X^+\rg\lg X^+|.
\end{equation}

	As a first framework or consistent family $\F_1$, we use the one
generated by $\alpha$, $\beta$ and $X$ at $t_1$, and $X^+$ and $X^-$ at
$t_2$. To
use the terminology of Sec~IV, $\F_1$ contains statements of the type
$p$: ``$S_x=1/2$ at time $t_1$'', $q$: ``the pointer at $t_2$
corresponds to the
apparatus having detected the particle emerging in the $S_x=1/2$ channel'', and
the like, which are then mapped to  projectors
\begin{equation}
	\varphi(p)= \alpha\otimes I,\ \ \varphi(q)=I\otimes X^+,
\end{equation}
etc., on $\H^2$.  The result is a Boolean algebra of projectors associated with
a family of simple histories at only two times, which is therefore
automatically consistent.

	Given the framework $\F_1$, it is possible to use the weights
(\ref{e4.w}) generated by the consistency functional, see \cite{gr84,gr93}, to
calculate various conditional probabilities such as:
\begin{equation}
	\Pr(\alpha,t_1\v X^+,t_2)=1,\ \ \Pr(\beta,t_1\v X^+,t_2)=0.
\end{equation}
Stated in words, given that the apparatus is in state $X^+$ at $t_2$,
one can be
sure that the particle was in the spin state $S_x=1/2$ and not $S_x=-1/2$ at
time $t_1$.  This is certainly the sort of result which one would expect to
emerge from a reasonable quantum theory of measurement, and which does, indeed,
emerge if one chooses an appropriate framework.

	Next, consider a situation in which for some reason one knows that the
particle at time $t_1$ has a spin polarization $S_z=1/2$
corresponding to the spin state
\begin{equation}
	|\gamma\rg = (|\alpha\rg + |\beta\rg)/\sqrt 2;
\label{e5b.5}
\end{equation}
for example, the particle may have come through a spin polarizer which selected
this polarization.  What will happen during the measurement process?  We adopt
as a framework the consistent family $\F_2$ which is generated by $\gamma$
($S_z=1/2$),
\begin{equation}
	|\delta\rg = (|\alpha\rg - |\beta\rg)/\sqrt 2
\end{equation}
($S_z=-1/2$), and $X$ at $t_1$; and, as before, $X^+$ and $X^-$ at $t_2$.
Using this framework (again, consistency is automatic) and the weights
generated by the consistency functional, one can calculate probabilities such
as:
\begin{equation}
	\Pr(X^+,t_2\v \gamma,X,t_1) = \Pr(X^-,t_2\v \gamma,X,t_1) = 1/2.
\label{e5b.7}
\end{equation}
Stated in words, given that $S_z=1/2$ and the apparatus was ready at $t_1$,
the probability is 1/2 that the apparatus is in the $X^+$ state, and 1/2 that
it is in the $X^-$ state at time $t_2$.

	All discussions within the framework of standard quantum mechanics
eventually arrive at the conclusion (\ref{e5b.7}), but many of them are forced
to make equivocations and unsatisfactory excuses along the way.  The reason is
that they adopt (implicitly) yet another framework, $\F_3$, which is generated
by $\gamma$ and $\delta$ at $t_1$, and at $t_2$ the state
\begin{equation}
	|G\rg = T(t_2,t_1)|\gamma,X\rg =
	(|\alpha',X^+\rg + |\beta',X^-\rg)/\sqrt 2.
\end{equation}
with projector $G$. To be sure, $\F_3$ is an acceptable framework, the
consistency conditions are satisfied, and within this framework one can derive
the conditional probability
\begin{equation}
	\Pr(G,t_2\v \gamma,X,t_1)=1.
\end{equation}
Stated in words, it is the case that {\it if one uses the consistent family
$\F_3$}, the state $G$ will occur with certainty at $t_2$, given that $S_z=1/2$
and the apparatus was ready at $t_1$.

	The problem, of course, is that $|G\rg$ is a macroscopic quantum
superposition (MQS), or Schr\"odinger's cat state, and tells one nothing
whatsoever about the position of the pointer on the apparatus.  Indeed,
adopting $\F_3$ makes it impossible even to discuss whether the apparatus is in
state $X^+$ or $X^-$ at the end of the measuring process, because the
projectors $X^+$ and $X^-$ do not commute with $G$.  All attempts to somehow
deduce that the pointer is in one position or the other ``for all practical
purposes'' are nonsensical once framework $\F_3$ has been adopted, in agreement
with Bell's observations \cite{bl90}, based upon more intuitive, but
nonetheless
quite reasonable arguments.  If one wants to talk about the pointer positions,
it is necessary to adopt a framework which includes the possibility of making
references to such positions, for example, $\F_2$.  But note that within $\F_2$
it is equally nonsensical to refer to the MQS state $G$.  Thus one sees that
the enormous confusion which surrounds most discussions of ``the measurement
problem'' is generated by a failure to distinguish the different frameworks or
consistent families which are possible in quantum theory, and to note that a
description employing one framework necessarily excludes certain statements
valid in other, incompatible frameworks.

	The consistent family $\F_2$, while it evades the problem of
``ghostly'' MQS states, can be faulted as a description of a {\it measurement}
process in that the outcome of the measurement in terms of pointer positions is
not correlated with a property of the particle before the measurement.  This
can be remedied by embedding $\F_2$ in a larger framework $\F_4$ which is
generated by the events already included in $\F_2$ at $t_1$ and $t_2$, and, in
addition, $\alpha$ and $\beta$, $S_x=\pm 1/2$, at a time $t_{1.5}$ which is
later than $t_1$ but earlier than the instant when the particle actually enters
the magnetic field gradient of the apparatus.  It turns out that $\F_4$ is
consistent (this must be checked, as it is no longer automatic), and using it
one can show, among other things, that
\begin{equation}
	\Pr(\alpha, t_{1.5}\v \gamma,X,t_1;X^+,t_2)=1,
\end{equation}
i.e., given the initial state at $t_1$, and the fact that at $t_2$ the pointer
indicates that the particle has emerged in the $S_x=1/2$ channel, it follows
that the spin state was $S_x=1/2$ at the time $t_{1.5}$.  Thus, using this
framework, one can again say that the apparatus after the measurement indicates
a state possessed by the particle before the measurement.  There is,
incidentally, nothing incompatible between having $S_z=1/2$ at $t_1$ and
$S_x=1/2$ at $t_{1.5}$ for the same particle, since these statements belong to
the same framework or consistent family.  This is just one of the ways in which
consistent quantum reasoning, with its precise rules, allows one to go well
beyond what is possible in standard quantum mechanics slavishly interpreted in
terms of ill-defined ``measurements''.

	This set of examples shows that a satisfactory theory of quantum
measurement requires the use of a framework which includes the possibility of
discussing {\it both} the outcome shown by the apparatus {\it and}, at a time
before the measurement, the properties of the measured system which the
measurement is designed to detect.  To be sure, precisely the same conditions
apply to a satisfactory theory of measurement in the context of classical
mechanics.  The difference is that in the classical case the choice of an
appropriate framework can be made implicitly with no difficulty, whereas in the
quantum case it is necessary to choose it with some care, in order to avoid
meaningless statements associated with attempts to mix descriptions belonging
to incompatible frameworks.

	\subsection{Spin of a Particle Between Two Measurements}
\label{between}

	The following example, with trivial differences in notation, is
from \cite{gr84}; applying the rules for descriptions contained in the present
paper clarifies the original presentation and responds to certain
criticisms \cite{des95,des87}.

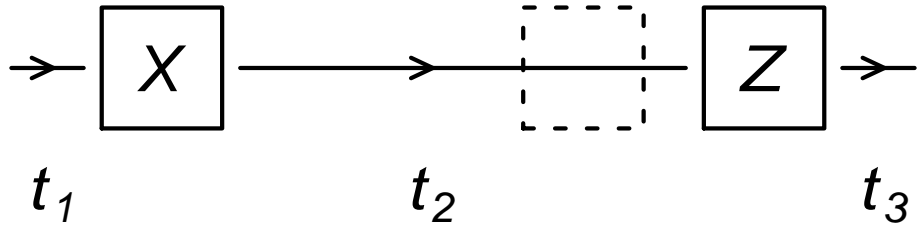
\begin{figure}[h]
\caption{Spin 1/2 particle passing through devices which measure $S_x$ and
$S_z$. A third device may be added at the position shown by the dashed line.}
\label{fig1}
\end{figure}

	Consider a spin 1/2 particle which passes through two successive
devices which measure the spin polarization, Fig.~1.  The first
device measures $S_x$ without changing it, so the unitary transformation is
\begin{equation}
	|\alpha,X\rg \ra |\alpha,X^+\rg,\ \
	|\beta,X\rg  \ra |\beta,X^-\rg,
\end{equation}
which is identical to (\ref{e5.m1}) except that the spin states $\alpha$
($S_x=1/2$) and $\beta$ ($S_x=-1/2$) are the same before and after the
measurement.  As in Sec.~\ref{measure}, $|X\rg$ is the ``ready'' state of the
device, and $|X^+\rg$ and $|X^-\rg$ correspond to the two different pointer
positions indicating the results of the measurement. The second device,
Fig.~\ref{fig1}, is similar, except that it measures $S_z$, with states $|Z\rg$
(``ready''), and $|Z^+\rg$ and $|Z^-\rg$ corresponding to having measured
$S_z=1/2$ and $S_z=-1/2$, respectively.

	As indicated in Fig.~\ref{fig1}, let $t_1$ be some time before the
particle enters the first device, when its spin state is $|\gamma\rg$,
corresponding to $S_z=1/2$; $t_2$ a time when it is between the two devices;
and $t_3$ a time when the particle has left the second device, and the pointers
on both devices indicate the results of the respective measurements.  Suppose
that at $t_3$ the measuring devices are in states $|X^+\rg$ and $|Z^+\rg$.
What can one conclude about the spin of the particle at the time $t_2$ when it
was between the two devices?

	First consider a consistent family $\F_1$ generated by the initial
state
\begin{equation}
	|\psi_1\rg=|\gamma XZ\rg
\end{equation}
at $t_1$ together with the pointer positions for both devices at $t_3$.
Consistency is automatic, as simple histories at only two times are involved.
Using this family, one can calculate the conditional probabilities for
the final
pointer positions given the initial state; for example,
\begin{equation}
	\Pr(X^+ Z^+, t_3 \v \psi_1,t_1)  = 1/4.
\label{e5.}
\end{equation}

	Next consider the family $\F_2$ obtained from $\F_1$ by adding $S_x$ at
time $t_2$.  This family is consistent, and using the corresponding weights
one can show that
\begin{equation}
	\Pr(S_x=1/2,t_2\v \psi_1,t_1; X^+ Z^+,t_3)=1;
\label{e5.bx}
\end{equation}
for the detailed calculation, see \cite{gr84}.  That is, given the
initial state
and the final pointer positions, one can be certain that at the intermediate
time $t_2$ the particle was in a spin state $S_x=1/2$.  Note that this
inference can be checked, in principle, by inserting a third device between the
first two, shown dashed in Fig.~1, which measures $S_x$, and verifying
that it yields the same result as the first device.

	One can also consider the family $\F_3$ obtained from $\F_1$ by adding
$S_z$ at time $t_2$.  This, too, is consistent, and with the help of the
corresponding weights (again, details are in \cite{gr84}) one can show that
\begin{equation}
	\Pr(S_z=1/2,t_2\v \psi_1,t_1; X^+ Z^+,t_3)=1.
\label{e5.bz}
\end{equation}
That is, given the same conditions as in (\ref{e5.bx}) one can conclude that at
the time $t_2$ the particle was in a spin state $S_z=1/2$.  This inference is
not particularly surprising when one remembers that, as shown in
Sec.~\ref{measure}, the outcome of a measurement allows one to infer the state
of the measured system prior to the time when the measurement takes place, if
one uses an appropriate consistent family.

	Note that there is no inconsistency between (\ref{e5.bx}) and
(\ref{e5.bz}), because the conclusion $S_x=1/2$ holds within framework $\F_2$,
and $S_z=1/2$ within framework $\F_3$, and these two frameworks are
incompatible with each other, even though each one separately is compatible
with $\F_1$. Thus we have an example of the possibility noted in
Sec.~\ref{s3b}, of two logical arguments based upon the same assumption, but
using incompatible frameworks, and whose conclusions cannot, therefore, be
combined.

	The correctness of $S_z=1/2$, within the framework $\F_3$, can be
verified, in principle, by inserting a third device between the first two,
Fig.~1, but this time of a type which measures $S_z$, and verifying that it
yields the same result as the final device the particle passes through.  One
may worry that this third device somehow ``creates'' a value of $S_z$ which was
``not there'' before the particle passed through it.  Such a worry is best put
to rest by means of a precise analysis, and the reader is invited to carry out
the appropriate calculations using a framework $\F_4$ in which, with all three
devices present, $S_z$ is specified both at $t_2$ and at a time $t_{2.5}$ when
the particle is between the two devices which measure $S_z$.  It is easily
shown that $S_z$ has the same value at both these times; thus, passing through
the intermediate device does not alter this component of the spin.

	\subsection{Double Slit}

\label{slit}

	A complete discussion of the paradoxes associated with double-slit
diffraction \cite{fls65} would require, at the very least, a theory of quantum
counterfactuals, and that is beyond the scope of the present article. A
significant insight into the source of the conceptual difficulties can,
nonetheless, be obtained using the tools of Secs.~III and IV\@.  In order to
focus on the essentials, we shall use the idealization that at time $t_1$ the
particle is described by a wavepacket $\psi_1({\bf r})$ which is approaching
the slits from the left in the geometry of Fig.~2, and that at a later time
$t_2$ the wave packet resulting from $\psi_1({\bf r})$ by unitary time
evolution consists of three pieces, two of which are waves confined to the
regions $R_A$ and $R_B$ just behind the two slits and moving to the right, and
the third a reflected wave located to the left of, and traveling away from the
slits.  And we shall regard the presence, at time $t_2$, of the particle in
$R_A$, represented by a projector $P_A$, as equivalent to the statement that
``the particle passed through slit $A$'', and likewise its presence in $R_B$,
projector $P_B$, as equivalent to its having passed through slit $B$.  In
addition, let the projector $P=P_A + P_B$ correspond to the region
\begin{equation}
	R=R_A\cup R_B.
\label{e5.sr}
\end{equation}
(The projector $P_A$ acting on a wave function $\psi({\bf r})$ produces a
function $\tilde\psi({\bf r})$ equal to $\psi({\bf r})$ for ${\bf r}$ inside
$R_A$, and equal to 0 for ${\bf r}$ outside $R_A$; $P_B$ is defined in
a similar way.)  By time $t_3$ the particle, if it has passed through the slit
system at all, will have been detected by one of the detectors in the
diffraction region, shown as circles in Fig.~2, where for convenience the
distance between the detectors and the slits has been considerably shortened.

	All the frameworks or consistent families in the discussion which
follows will include an initial state
\begin{equation}
	|\Psi_1\rg = |\psi_1, D_1, D_2,\ldots D_m\rg
\end{equation}
at time $t_1$, where $D_j$ indicates that detector $j$ is ready to detect a
particle. The first family $\F_1$ we wish to consider has, in addition to the
initial state, projectors $D_1^*, D_2^*,\ldots$ at $t_3$, where $D_j^*$ is a
projector indicating that detector $j$ has detected a particle.  Since $\F_1$
involves simple histories at just two times, the consistency condition is
satisfied, and the weights (\ref{e4.w}) may be used to calculate the
probability
\begin{equation}
	\Pr(D_j^*, t_3\v \Psi_1,t_1)
\label{e5.d}
\end{equation}
that, given the initial conditions, the $j'$th counter will trigger.  Its
dependence on $j$ will exhibit the usual diffraction pattern.

	The next family of interest, $\F_2$, is obtained by adding to the
events of $\F_1$ an event corresponding to the particle being in the region $R$
(\ref{e5.sr}), projector $P$, at time $t_2$.  This is again a consistent family
in which one can calculate quantities such as
\begin{equation}
	\Pr(P,t_2\v \Psi_1,t_1)
\end{equation}
the probability that the particle was not reflected by the slit system, and
\begin{equation}
	\Pr(D_j^*,t_3\v P,t_2;\Psi_1,t_1),
\end{equation}
the probability that the $j'$th counter will trigger if the particle passes
through the slit system.  The latter shows the same diffraction pattern as
(\ref{e5.d}), as a function of $j$, aside from a multiplicative constant.  On
the other hand, attempting to add to $\F_2$ the events that the particle is in
one of the regions $R_A$ or $R_B$ at $t_2$ results in an inconsistent family,
despite the fact that the corresponding projectors $P_A$ and $P_B$ can be
included in a Boolean algebra of projectors for the tensor product $\H^3$.
Thus, while one can consistently say that ``the particle was in $R$'' at time
$t_2$, it makes no sense, given this framework, to say that ``the particle was
in $R_A$, or it was in $R_B$''.  But given that $R$, (\ref{e5.sr}), is the
union of two disjoint regions $R_A$ and $R_B$, it is difficult not to interpret
the second phrase as equivalent to the first. Thus we have a situation where it
may actually be helpful, in order to avoid confusion, to employ the projectors
themselves in place of the corresponding English phrases.

	The preceding discussion shows that the consistent history analysis
supports the usual intuition that it is somehow improper to talk about which
slit the particle passed through.  However, it replaces an intuition which is
always a little vague with a precise mathematical criterion based upon the
Hilbert space structure of quantum mechanics.  That this represents a
significant advance is evident when one comes upon circumstances, such as that
in which a detector is placed directly behind one of the slits, in which it
seems intuitively plausible that one should be able to specify the slit through
which the particle passed.

	Even without displacing the detectors, one can construct a framework
which permits one to say which slit the particle passed through, once again for
convenience interpreting this as an assertion about its presence in $R_A$ or
$R_B$ at $t_2$.  Let $\F_3$ be the family generated by $P_A$ and $P_B$ at $t_2$
along with $\Psi_1$ at $t_1$, and (for the moment) no events at $t_3$. As these
simple histories involve only two times, consistency is automatic.  One can
then calculate the probabilities
\begin{equation}
	\Pr(P_A,t_2\v \Psi_1,t_1),\ \Pr(P_B,t_2\v \Psi_1,t_1),
\end{equation}
that the particle passed through slit $A$ and slit $B$, respectively, and
demonstrate that the particle surely did {\it not} pass through {\it both}
slits by using the fact that $P_AP_B=\emptyset$. If, however, one
attempts to add
to $\F_3$ events of the sort $D_j^*$ at $t_3$, the result is an inconsistent
family.  From this we see that what prevents one from sensibly talking about
which slit the particle passes through in the usual discussions of double slit
diffraction is directly tied to the (implicit) requirement of being able to
describe the point at which the particle will arrive, or be detected, in the
diffraction plane.  A description which includes which slit the particle goes
through can also consistently include events referring to the detectors at time
$t_3$; these events must, however, be suitable MQS states.  This suggests that
the absence of a precise, clear discussion of two-slit diffraction in textbooks
is not unrelated to the absence of a clear treatment of MQS states and the
measurement problem. A consistent logical analysis using the concept of a
framework is able to dispose of both problems simultaneously.

	\subsection{Emergence of the Classical World}

\label{world}

	Both Gell-Mann and Hartle \cite{gmh}, and Omn\`es \cite{om} have
discussed, from slightly different perspectives, how classical physics
expressed in terms of suitable ``hydrodynamic'' variables emerges as an
approximation to a fully quantum-mechanical description of the world when the
latter is carried out using suitable frameworks, i.e., families of consistent
histories.  It is not our purpose to recapitulate or even summarize their
detailed technical discussions, but instead to indicate the overall strategy,
as viewed from the perspective of this paper, and how it is related to
processes of ``decoherence''  \cite{decohere} which arise when a particular
quantum subsystem of interest interacts with a suitable environment,
both of which are part of a larger (closed) quantum system.

	The basic strategy of Gell-Mann and Hartle can be thought of as the
search for a suitable ``quasi-classical'' framework, a consistent family whose
Boolean algebra includes projectors appropriate for representing coarse-grained
variables, such as average density and average momentum inside volume elements
which are not too small, variables which can plausibly be thought of as the
quantum counterparts of properties which enter into hydrodynamic and other
descriptions of the world provided by classical physics.  Hence it is necessary
first to find suitable commuting projectors, and then to show that the
consistency conditions are satisfied for the corresponding Boolean algebra.
Omn\`es states his strategy in somewhat different terms which, however, seem at
least roughly compatible with the point of view just expressed.  (The actual
technical calculations of Gell-Mann and Hartle, and of Omn\`es are based on
simple, rather than generalized histories, in the terminology of
Sec.~\ref{s4b}.)

	One might worry that the strategies of Omn\`es, and Gell-Mann and
Hartle, are incompatible with the formulation of quantum theory contained in
the present paper, because their consistency conditions employ a density
matrix, whereas that in Sec.~\ref{s4c} does not.  However, the difference is
probably of no great importance when discussing ``quasi-classical'' systems
involving large numbers of particles, for the following reason.  In classical
statistical mechanics one knows (or at least believes!) that for macroscopic
systems the choice of ensemble---microcanonical, canonical, or grand---is for
many purposes unimportant, and, indeed, the average behavior of the ensemble
will be quite close to that of a ``typical'' member. Stated in other words, the
use of probability distributions is a convenience which is not ``in principle''
necessary.  Presumably an analogous result holds for quantum systems of
macroscopic size: the use of a density matrix, both as an ``initial condition''
and as part of the consistency requirement may be convenient, but it is not
necessary when one is discussing the behavior of a closed system.  Of course,
there are issues here which deserve serious study; the proper formulation of
time-dependent phenomena in quantum statistical mechanics is an open problem,
Sec.~\ref{s6b}.

	The task of finding an appropriate quasi-classical consistent family is
made somewhat easier by two facts.  The first is that decoherence (as defined
above) is quite effective in reducing ``off-diagonal'' terms $C(\tilde
M^{(\alpha)},\tilde M^{(\beta)})$, with $\alpha\neq\beta$, in the consistency
functional (\ref{e4.cf}) for a suitably chosen Boolean algebra representing
quasi-classical variables in circumstances in which thermodynamic
irreversibility plays a role.  The second is that the consistency functional is
a continuous function of its arguments, and hence it is plausible that by
making small changes in the projectors forming the Boolean algebra, one can
reduce the off-diagonal elements to zero, assuming they are already small (see
Dowker and Kent \cite{dk95}) so that the consistency condition (\ref{e4.c}) is
satisfied.  Since there is in any case some arbitrariness in choosing which
quantum projectors to associate with particular coarse-grained hydrodynamic
variables, small changes in these projectors are unimportant in terms of their
physical interpretation.  Thus exact consistency does not seem difficult to
achieve ``in principle'', even if in practice theoretical physicists are
unlikely to be worried if the ``off-diagonal'' elements of the consistency
functional are not precisely zero, as long as they are suitably small in
comparison with the ``diagonal'' weights $C(\tilde P,\tilde P)$ which enter
into a calculation of probabilities.

	There are many frameworks which are {\it not} quasi-classical, and from
a fundamental point of view there is no reason to exclude using one of these to
describe the world.  The fundamental principles of quantum mechanics no more
dictate the choice of a framework than the concepts of geometry dictate which
projection a draftsman must use for representing a three-dimensional object.
In both cases the issues are practical ones, related to what one is trying to
achieve. As noted in Sec.\ref{measure}, a framework which includes MQS states
of a measurement apparatus after the measurement precludes any attempt to think
of the process of its interaction with the measured system as a ``measurement''
in the ordinary sense of that word.  Using MQS states to describe
Schr\"odinger's cat does nothing whatsoever to the cat, but it does make it
impossible to discuss whether the cat is or is not alive, for such a discussion
will employ concepts of a quasi-classical kind which cannot be included in the
chosen framework.  And any quantum description which wishes to make contact
with the world of everyday experience or of experimental physics must employ
some framework which allows a description of the appropriate kind.

	As a final remark, the fact that in Sec.~\ref{measure} above no mention
was made of decoherence in addressing the ``measurement problem'' should not be
taken to mean that considerations of decoherence are irrelevant to discussions
of quantum measurements; quite the opposite is the case.  For example, the fact
that certain physical properties, such as pointer positions in a properly
designed apparatus, have a certain stability in the course of time despite
perturbations from a random environment, while other physical properties do
not, is a matter of both theoretical and practical interest.  However,
decoherence by itself cannot single out a particular framework of consistent
histories, nor can it disentangle conceptual dilemmas brought about by mixing
descriptions from incompatible frameworks.  To the extent that the latter form
the heart of the ``measurement problem'', decoherence will not resolve it, and
claims to the contrary merely add to the confusion.



		\section{Summary and Open Questions}
\label{s6}

	\subsection{Summary}
\label{s6a}

	The description of a quantum system and the reasoning process relating
one description to another are best thought of in terms of a {\it framework},
which serves to establish, in effect, a topic of conversation or a point of
view, and {\it statements}, which are assertions about the state of a quantum
mechanical system which are either true or false within the framework with
which they are associated.  A classical analogy, Sec.~\ref{s2a}, in which a
framework must be chosen by the person constructing the description, and truth
is relative to this framework, suggests that there is no reason to treat
quantum descriptions as less ``objective'' than their classical counterparts.
The most important difference between classical and quantum descriptions is the
fact that in the quantum case there are {\it incompatible} frameworks, any one
of which could be employed for constructing descriptions of a given system, but
which cannot be combined, either for the process of describing the system or
reasoning about it. This incompatibility has no direct classical analog, and it
leads to the rule which states that any description of a quantum system must
either employ a single framework or a compatible family of frameworks.
Similarly, a logical argument, leading from certain descriptions, taken as
assumptions, to other descriptions which form the conclusions, must use a
single framework or a compatible family of frameworks.  In the case of a
classical mechanical system, a single framework suffices for all descriptions,
and this framework is often chosen implicitly.  A quantum framework can also be
chosen implicitly, but carelessness can lead to invalid reasoning and to
paradoxes.

	The same combination of framework and statement can be used to describe
the behavior of a quantum system as a function of time.  In this case the
framework is usually called a ``consistent family of histories''.  Once again,
a valid description or history must be based upon a single framework or
consistent family, or a compatible set of consistent families, and a logical
argument leading from assumptions to conclusions must likewise employ a single
consistent family or a compatible set of families.  It is convenient to
represent (generalized) histories involving $n$ times using projectors on the
tensor product of $n$ copies of the Hilbert space.  The consistency condition
can then be expressed using a bilinear {\it consistency functional\/} of such
projectors, and when consistency is satisfied, the weights which determine
conditional probabilities can also be computed using this consistency
functional.

	Applications of the formalism just described include measuring
processes, the state of a quantum system between measurements, double-slit
diffraction, and the emergence of the classical world from a fully quantum
mechanical theory.

	\subsection{Open Questions}
\label{s6b}

	 The discussion of consistency in Sec.~\ref{s4c} was carried out
assuming a finite-dimensional Hilbert space in order to ensure, among other
things, that the trace (\ref{e4.cf}) and the minimal elements (\ref{e4.di}) of
the Boolean algebra $\BT$ exist. It seems likely that by making suitable
restrictions on $\BT$, the concept of consistency can be extended in a
satisfactory way to an infinite-dimensional Hilbert space, but this needs to
be worked out in detail.  There are similar technical issues associated with
frameworks which employ an infinite collection of elementary statements, the
compatibility of infinite collections of frameworks, and histories defined at
an infinite set of times.

	The structure for quantum reasoning proposed in this paper includes no
mechanism for dealing with counterfactuals (``if the counter had not been
located directly behind the slit, then the particle would have $\ldots$'').
Inasmuch as many quantum paradoxes, including some of the ones associated with
double-slit diffraction, and certain derivations of Bell's inequality and
analogous results, make use of counterfactuals, disentangling them requires an
analysis which goes beyond what is found in the present paper.  As philosophers
have yet to reach general agreement on a satisfactory scheme for counterfactual
reasoning applied to the classical world \cite{rhpc}, an extension which covers
all of quantum reasoning is likely to be difficult. On the other hand, a scheme
sufficient to handle the special sorts of counterfactual reasoning found in
common quantum paradoxes probably represents a simpler problem.

	The proper formulation of time-dependent phenomena in quantum
statistical mechanics remains an open question.  On the one hand, as noted in
Sec.~\ref{world}, it seems plausible that various formulas which yield the
``average'' behavior of some macroscopic system are not in need of revision,
since, among other things, this ``average'' is likely to be essentially the
same as the behavior of a typical member of an ensemble, and because
decoherence (interaction with the environment) is very effective in removing
violations of the consistency conditions if one chooses an appropriate
``quasi-classical'' consistent family.  The results obtained in  \cite{om,gmh}
are quite encouraging in this connection.  On the other hand, thought needs to
be given to those cases in which different members of an ensemble have
significantly different behavior, that is, where ``fluctuations'' are
important.  In any event, it would be useful to have a restatement of the
fundamental principles of quantum statistical mechanics based upon a fully
consistent interpretation of standard quantum mechanics.

	Can the structure of reasoning developed in this paper for
non-relativistic quantum mechanics be extended to relativistic quantum
mechanics and quantum field theory?  Various examples suggest that the sort of
peculiar non-locality which is often thought to arise from violations
of Bell's inequality and various EPR paradoxes will disappear when one enforces
the rule that consistent quantum arguments must employ a single framework.
While this is encouraging, it is also true that locality (or the lack thereof)
in non-relativistic quantum theory has not yet been carefully analyzed from the
perspective presented in this paper, and hence must be considered among
the open
questions.  And, of course, getting rid of spurious non-localities is only a
small step along the way towards a fully relativistic theory.

	Even in the domain of non-relativistic (and non-counterfactual) quantum
reasoning, there is no proof that the scheme presented in this paper is
appropriate and adequate for all of standard quantum mechanics.  Indeed,
standard quantum mechanics contains a great deal of seat-of-the-pants intuition
which has never been formalized, and thus it is hard to think of any way of
testing the system of reasoning presented here apart from applying it to a
large number of examples, to see if it yields what the experts agree are the
``right answers'', and what the experimentalists find in their laboratories.
During the past ten years consistent history ideas have been applied to many
different situations without encountering serious problems, but there is always
the possibility that the fatal flaw lies just around the next corner.  If such
a flaw exists, it should not be hard to identify, since the scheme of reasoning
presented in this paper has precise rules and is based on ``straight'' quantum
mechanics, without hidden variables, adjustable parameters, modifications of
the Schr\"odinger equation, and excuses of the ``for all practical purposes''
type.  The reader who considers it defective is invited to point out the
problems!



\section*{Acknowledgements}

	It is a pleasure to acknowledge stimulating correspondence and/or
conversations with B. d'Espagnat, F. Dowker, L. Hardy, J. Hartle, A.
Kent, R.  Omn\`es, M. Redhead, and E. Squires.  Financial support for this
research has been provided by the National Science Foundation through grant
PHY-9220726.





\section*{Figure Captions}

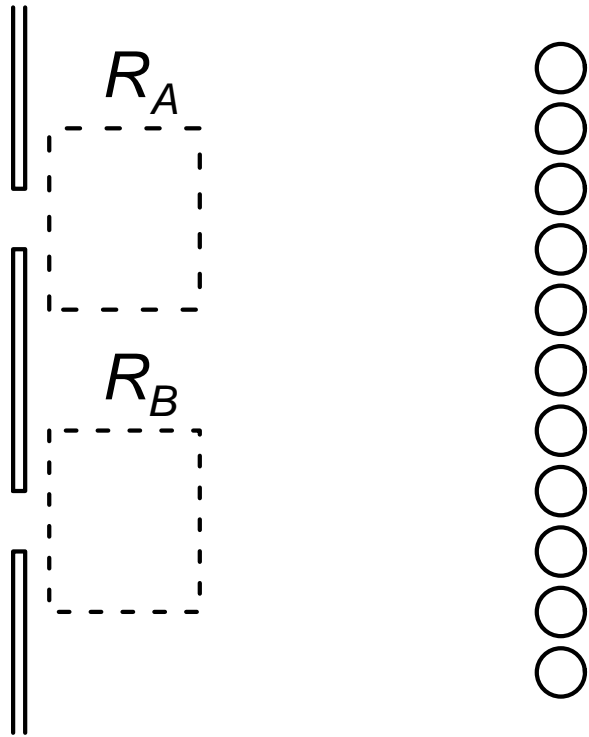
\begin{figure}[h]
\caption{Diffraction of a particle by two slits (see text).}
\label{fig2}
\end{figure}


\begin{thebibliography}{11}
\bibitem{fn64}

R.~Feynman {\it The Character of Physical Law} (M.I.T. Press, Cambridge,
Mass., 1965) p.~129.

\bibitem{bl64}

J. S. Bell, Physics {\bf 1}, 195 (1964); reprinted in J. A. Wheeler and
W. H. Zurek
(editors),{\it Quantum Theory and Measurement} (Princeton University Press,
Princeton, 1983).

\bibitem{lit1}
	The following are but a few drops in the ocean of literature on this
subject: 	F. Selleri (editor), {\it Quantum Mechanics Versus Local
Realism} (Plenum Press, New York, 1988); 	M. Redhead, {\it
Incompleteness, Nonlocality, and Realism} (Oxford University Press, Oxford,
1987), Ch. 4; 	N. D. Mermin, Rev. Mod. Phys. {\bf 65}, 803 (1993).

\bibitem{ess92}
	B. G. Englert, M. O. Scully, G. Sussmann and H. Walther, Z.
Naturforsch. {\bf 47a}, 1175 (1992); {\bf 48a}, 1261 (1993)

\bibitem{bom52}

	D. Bohm, Phys. Rev. {\bf 85}, 166 (1952); {\bf 85}, 180 (1952). These
are reprinted in J. A. Wheeler and W. H. Zurek
(editors),{\it Quantum Theory and Measurement} (Princeton University Press,
Princeton, 1983).

\bibitem{bhi93}
	D. Bohm and B. J. Hiley, {\it The Undivided Universe} (Routledge,
London, 1993).

\bibitem{dgz92}
	For example,
D. D\"urr, S. Goldstein and N. Zangh\'\i, J. Stat. Phys. {\bf 67}, 843  (1992).


\bibitem{dhs93}
	C. Dewdney, L. Hardy and E. J. Squires, Phys. Lett. A {\bf 184}, 6
(1993).

\bibitem{dfg93}

D. D\"urr, W. Fusseder, S. Goldstein and N. Zanghi, Z. Naturforsch. {\bf 48a},
1261 (1993).

\bibitem{vn55a}
J. von Neumann {\it Mathematical Foundations of Quantum Mechanics}
(Princeton University Press, Princeton, 1955), Ch. III,
Sec. 5.

\bibitem{vn55b}
Ibid., Ch. VI


\bibitem{bvn36}
G. Birkhoff and J. von Neumann,
Annals of Math. {\bf 37}, 823 (1936).

\bibitem{gr84}
R. B. Griffiths, J. Stat. Phys. {\bf 36}, 219  (1984).

\bibitem{om}

R. Omn\`es, J. Stat. Phys. {\bf 53}, 893 (1988); Rev. Mod.  Phys. {\bf 64}, 339
(1992);  {\it The Interpretation of Quantum Mechanics} (Princeton
University Press, Princeton, 1994).

\bibitem{gmh}

M. Gell-Mann and J. B. Hartle in {\em Complexity, Entropy, and the Physics of
Information}, edited by W. Zurek (Addison Wesley, Reading, 1990); Phys. Rev. D
{\bf 47}, 3345 (1993).




\bibitem{logic}

	Yu. I. Manin, {\it A Course in Mathematical Logic} (Springer-Verlag, New
York, 1977);
	E. Mendelson, {\it Introduction to Mathematical Logic}, 3d ed.
(Wadsworth, Inc., Belmont, Calif., 1987).


\bibitem{pz93}
J. P. Paz and W. H. Zurek, Phys. Rev. D {\bf 48}, 2728 (1993).

\bibitem{des95}
B. d'Espagnat, {\it Veiled Reality} (Addison-Wesley, Reading,
Mass., 1995), Sec. 11.4.

\bibitem{dk95}
F. Dowker and A. Kent, submitted to J. Stat. Phys.

\bibitem{ompc}
R. Omn\`es, private communication.

\bibitem{gr94}

	R. B. Griffiths, in {\it Symposium on the Foundations of Modern Physics
1994}, edited by  K.~V.~Laurikainen, C.~Montonen, and K.~Sunnarborg (Editions
Fronti\`eres, Gif-sur-Yvette, Framce, 1994), p. 85.

\bibitem{gr93}
R. B. Griffiths, Phys. Rev. Lett. {\bf 70}, 2201  (1993).

\bibitem{gr93b}

R. B. Griffiths, Found. Phys. {\bf 23}, 1601 (1993).



\bibitem{bl90} 
J. S. Bell, in {\it Sixty-Two Years of
Uncertainty}, edited by A. I. Miller (Plenum Press, New York, 1990) p. 17.

\bibitem{des87}

B. d'Espagnat,
Phys. Lett. A {\bf 124}, 204 (1987).

\bibitem{fls65}
	R. P. Feynman, R. B. Leighton, and M. Sands, {\it The Feynman Lectures
on Physics} (Addsion-Wesley, Reading, Mass., 1965), Vol. III, Ch. 1.

\bibitem{rhpc}

	M. Redhead, private communication.

\bibitem{decohere}

	See \cite{pz93} and references given there.

\end{thebibliography}
\end{document}